\documentclass[a4paper,12pt]{article}
\usepackage{tikz}
\usepackage{simplewick}
\usepackage{jheppub}

\title{Mandelstam formulae for
generalised Wilson loops} 
\author{Chris Curry and Paul Mansfield \\Centre for Particle Theory, University of Durham, Durham DH1 3LE, UK \\ Email: c.h.curry@durham.ac.uk, p.r.w.mansfield@durham.ac.uk}
\abstract{
We derive Mandelstam formulae for two generalisations of the Wilson loop. In these generalisations path-ordering of Lie algebra generators is replaced by an anti-commuting one dimensional field theory along the loop. 
We extend the calculation to the $N=1$ super-Wilson loop by introducing a superpartner for the additional field.}


\preprint{}

\begin{document}

\maketitle

\section{Introduction}

The non-Abelian Wilson loop is the trace of the path-ordered exponential of a non-Abelian gauge field integrated along a closed curve, $C$,  in spacetime, given parametrically as $x^\mu=x^\mu(\xi)$,
\begin{equation}
W\equiv \text{tr} \bigg( \mathcal{P} \ \text{exp}\bigg(\oint_C d\xi\,A_{\mu}\dot x^{\mu}\bigg)\bigg)\\.\label{one}
\end{equation}
It
is an important observable in quantum gauge theory. For instance, local observables can be written in terms of Wilson loops. They are also useful in studying the confining properties of the gauge theory since they can distinguish the confinement phase by the so called area law \cite{Wilson:1974sk}.  Another use of Wilson loops is to provide the coupling of particles to gauge-fields in first quantisation, for example the partition function of a spin-0 particle of mass $m$ coupled to the gauge field is
\begin{equation}
 \int {\cal D}(x,\sqrt{h})\,W[C]\,\exp \left(-\frac{1}{2}\int d\xi \,\left(\frac{\dot x\cdot\dot x}{\sqrt{h}}
 +m^2\sqrt{h}\right)\right)\,.\label{spin0}
\end{equation}
We can expand the gauge field as $A_{\mu}^A\tau^A$, where $\tau^A$ are anti-Hermititan Lie algebra generators. It is well known that if $\psi^\dagger_r$ and $\psi_s$ are a set of anti-commuting operators with $\{\psi^\dagger_r,\,\psi_s\}=\delta_{rs}$ then the operators $\hat \tau^R\equiv\psi^\dagger_r\, \tau^R_{rs}\,\psi_s$ satisfy the Lie algebra. These anti-commutation relations follow from a Lagrangian $\psi^\dagger\dot\psi$, which leads to a propagator containing the step-function $\theta(t_1-t_2)$ (plus other terms depending on the boundary conditions) which is just what is needed to build the path-ordering in (\ref{one})\cite{Samuel:1978iy}. So instead of (\ref{one}) we might consider
\begin{equation}
W_{\psi}\equiv \int \mathcal{D}[\psi^{\dagger},\psi] \ \text{exp}\bigg(\oint_C d\xi \ \psi^{\dagger}(\dot{\psi}+\dot{x}^{\mu}A_{\mu}\psi)\bigg). \label{psi loop 2}
\end{equation}
The Lagrangian in this model has been analysed extensively \cite{Bala}-\cite{Salomonson:1977as}, mainly in canonical quantisation. The path-ordered exponential in (\ref{one}) can be picked out by a particular choice of boundary conditions, an operator insertion and a choice of normalisation, or, as in \cite{D'Hoker:1995bj}
and \cite{Bastianelli:2013pta}, by projecting out a piece with appropriate $U(1)$-charge. We shall not follow these paths, but rather we will find it more useful to consider the integral (\ref{psi loop 2}) as it stands. Integrating out the $\psi$-field (with some choice of boundary conditions) gives \cite{octopus}
\[
 W_\psi={\rm Det}\left( \frac{d}{d\xi}+\dot{x}^{\mu}A_{\mu}\right)
\]
\[
 \propto{\rm det}\left(\sqrt{e^{-\oint_C A\cdot dx}}\,\pm\, 1/\sqrt{e^{-\oint_C A\cdot dx}}\right)\,,
\]
where the signs correspond to anti-periodic and periodic $\psi$.
This is then a generalisation of the Wilson-loop but is one that is closely related and has useful properties. In \cite{octopus} it was shown that when $W[C]$ is replaced by $W_\psi$ in the spin-1/2 generalisation of (\ref{spin0}) which introduces  fermionic superpartners of $x$ to describe $\gamma$-matrices then the partition function picks out the representations and helicities of the hadrons and leptons that appear in a single generation of the Standard Model when the same boundary conditions are imposed on all the fermionic variables. This suggests that $W_\psi$ is worth studying in its own right, so in this paper we will construct the loop equations for this quantity generalising those constructed for the Wilson loop itself.

There is another, more speculative reason for studying $W_\psi$ rather than $W[C]$ related to attempts to represent the gauge field dynamics in terms of lines of force spanning $C$. When averaged over $A$, using
\[
 \langle\,\Omega\,\rangle\equiv\frac{1}{Z_0}\int \mathcal{D}A  \ e^{-S_{YM}}\,\Omega\,,
\]
where
\begin{equation}
S_{YM}=\frac{1}{4q^2}\int d^4x \ F^{\mu\nu A}F_{\mu\nu}^A \label{maxwell}
\end{equation}
\begin{equation}
F_{\mu\nu}=\partial_{\mu}A_{\nu}-\partial_{\nu}A_{\mu}+[A_{\mu},A_{\nu}]
\end{equation}
\begin{equation}
Z_0=\int \mathcal{D}A \ e^{-S_{YM}}
\end{equation}
the leading perturbative contribution comes from the free part of the Yang-Mills action
\[\langle W_\psi\rangle=
\]
\begin{equation}
 \int \mathcal{D}[\psi^{\dagger},\psi]  \text{exp}\bigg(-S_{kin}- 
 \frac{1}{2}\int \frac{d^4k}{(2\pi)^4}\oint\oint \ dx^{\mu}\psi^{\dagger}\tau^A\psi|_{\xi} \  \frac{e^{ik\cdot(x(\xi)-x(\xi'))}}{k^2} \ dx_{\mu}\psi^{\dagger}\tau^A\psi|_{\xi'}\bigg). \label{wilson psi}
\end{equation}
Apart from the $\psi$-dependence this is the Abelian result. In \cite{Edwards:2014xfa} it was shown how to reproduce the Abelian result from a tensionless string with non-standard interaction whose world-sheet spans $C$ . The appearance of $\psi^\dagger$ and $\psi$ in the perturbative expansion of the Yang-Mills case suggests that these extra variables might be boundary values of world-sheet fields that appear when we try to generalise to the non-Abelian case in which event they should  be responsible for the self-interactions of Yang-Mills theory. This was explored in \cite{Curry} with only limited success as unfortunately, the model studied there lacks the singularity structure required to incorporate the self interactions, though path-ordering is achieved. Thus, a further extension of this model is required to complete the reformulation. In the present paper we shall not explore this idea further but simply study the expectation value of $W_\psi$ when the gauge field dynamics are the standard ones with the 
hope that the resulting loop equations may provide a useful tool in searching for the correct string theory model.

The loop equations are functional equations for $\langle W[C]\rangle$ as $C$ is varied. They have been much investigated \cite{Gervais:1978mp}- \cite{Makeenko:1980vm}. Here we will follow the general approach of \cite{Polyakov:1987ez} to work out the corresponding equations for $W_\psi$ and its supersymmetric extension. This approach is based on
the Mandelstam formula for the Wilson loop:
\begin{equation}
\Delta(\xi) \langle \,W[C]\,\rangle=-q^2 \ \mathcal{P} \ \oint \ \dot{x}^{\mu}\tau^A|_{\xi} \ \delta^4(x(\xi)-x(\xi')) \ dx_{\mu}\tau^A|_{\xi'} \langle e^{-\oint_C A\cdot dx}\rangle \label{mandelstam}
\end{equation}
where 
\begin{equation}
\Delta(\xi)\equiv \lim_{\epsilon\rightarrow 0}\int^{\epsilon}_{-\epsilon} d\xi' \ \frac{\delta^2}{\delta x^{\mu}(\xi+\xi'/2)\delta x_{\mu}(\xi-\xi'/2)}  \label{area}
\end{equation}
is the Laplacian in loop space \cite{Polyakov:1987ez}.

\section{Bosonic Theory}

We begin by considering the change in $W_{\psi}$ under a variation of $C$
\begin{equation}
\delta_xW_{\psi}=\int \mathcal{D}[\psi^{\dagger},\psi] \ e^{\oint d\xi \  \psi^{\dagger}(\dot{\psi}+q\dot{x}^{\mu}A_{\mu}\psi)} \oint d\xi \bigg(\psi^{\dagger}\delta\dot{x}^{\mu}A_{\mu}\psi+\psi^{\dagger}\dot{x}^{\nu}\delta x^{\mu}\partial_{\mu}A_{\nu}\psi\bigg).
\end{equation}
After an integration by parts on the first term this becomes
\begin{equation}
\delta_xW_{\psi}=\int \mathcal{D}[\psi^{\dagger},\psi] \ e^{\oint d\xi \  \psi^{\dagger}(\dot{\psi}+\dot{x}^{\mu}A_{\mu}\psi)}\oint d\xi \bigg(\psi^{\dagger}\delta x^{\mu}\dot{x}^{\nu}(\partial_{\mu}A_{\nu}-\partial_{\nu}A_{\mu})\psi \nonumber
\end{equation}
\begin{equation}
-\delta x^{\mu}\big(\dot{\psi}^{\dagger}A_{\mu}\psi+\psi^{\dagger}A_{\mu}\dot{\psi}\big)\bigg). \label{var W}
\end{equation}
The second line can be replaced by considering the Schwinger-Dyson equations for $\psi^{\dagger}$ and $\psi$.
Under a variation of $\psi^{\dagger}$ (and $\psi$), the functional $W_{\psi}$ does not change, providing the functional measure $\mathcal{D}\psi^{\dagger}$ (and $\mathcal{D}\psi$) doesn't change. Therefore, we obtain the relations
\begin{equation}
0=\delta_{\psi^{\dagger}}W_{\psi}=\int \mathcal{D}[\psi^{\dagger},\psi] e^{\oint d\xi \  \psi^{\dagger}(\dot{\psi}+\dot{x}^{\mu}A_{\mu}\psi)}\oint d\xi \ \delta\psi^{\dagger}(\dot{\psi}+\dot{x}^{\mu}A_{\mu}\psi)
\end{equation}
\begin{equation}
0=\delta_{\psi}W_{\psi}=\int \mathcal{D}[\psi^{\dagger},\psi] e^{\oint d\xi \  \psi^{\dagger}(\dot{\psi}+\dot{x}^{\mu}A_{\mu}\psi)}\oint d\xi \ (-\dot{\psi^{\dagger}}+\psi^{\dagger}\dot{x}^{\mu}A_{\mu})\delta\psi.
\end{equation}
Choosing the specific variations $\delta\psi^{\dagger}=\psi^{\dagger}\delta x^{\mu}A_{\mu}$ and $\delta \psi=\delta x^{\mu}A_{\mu}\psi$ allows us to replace the time derivative terms of (\ref{var W}) with a commutator, completing the appearance of the field strength in the variation of $W_{\psi}$. This choice also ensures that the Jacobian arising from a change of variables is trivial. With these relations, the change in $W_{\psi}$ can be written as
\begin{equation}
\delta_xW_{\psi}=\int \mathcal{D}[\psi^{\dagger},\psi] \ e^{\oint d\xi \  \psi^{\dagger}(\dot{\psi}+\dot{x}^{\mu}A_{\mu}\psi)}\oint d\xi \  \psi^{\dagger}\delta x^{\mu}\dot{x}^{\nu}F_{\mu\nu}\psi. \label{first var}
\end{equation}
Varying the loop variable for a second time we find
\begin{equation}
\delta_2\delta_1W_{\psi}=\int \mathcal{D}[\psi^{\dagger},\psi] \ e^{\oint d\xi \  \psi^{\dagger}(\dot{\psi}+\dot{x}^{\mu}A_{\mu}\psi)}\bigg(\oint d\xi \ \psi^{\dagger}(\delta_1 x^{\mu}\delta_2 \dot{x}^{\nu}F_{\mu\nu}+\delta_1 x^{\mu} \delta_2 x^{\alpha}\dot{x}^{\nu}\partial_{\alpha}F_{\mu\nu})\psi \nonumber
\end{equation}
\begin{equation}
+\oint d\xi' \ \psi^{\dagger}\dot{x}^{\beta}\delta_2 x^{\alpha}(\partial_{\alpha}A_{\beta}-\partial_{\beta}A_{\alpha})\psi     \oint d\xi \ \psi^{\dagger}\delta_1 x^{\mu}\dot{x}^{\nu}F_{\mu\nu}\psi \nonumber
\end{equation}
\begin{equation}
-\oint d\xi' \ (\dot{\psi}^{\dagger}\delta_2 x^{\alpha}A_{\alpha}\psi+\psi^{\dagger}\delta_2 x^{\alpha}A_{\alpha}\dot{\psi})\oint d\xi \ \psi^{\dagger}\delta_1 x^{\mu}\dot{x}^{\nu}F_{\mu\nu}\psi \bigg) \,
\end{equation}
after an integration by parts.
We may invoke the Schwinger-Dyson equations for (\ref{first var}) again for the variables $\psi^{\dagger}$ and $\psi$. Choosing the same specific variations for $\delta\psi^{\dagger}$ and $\delta\psi$ as above, these give the relations
\begin{equation}
\int \mathcal{D}[\psi^{\dagger},\psi] \ e^{\oint d\xi \  \psi^{\dagger}(\dot{\psi}+\dot{x}^{\mu}A_{\mu}\psi)} \oint d\xi \ \delta_1 x^{\mu}\dot{x}^{\nu}\psi^{\dagger}F_{\mu\nu}\psi\oint d\xi' \ \delta_2 x^{\alpha}\psi^{\dagger}A_{\alpha}\dot{\psi}   \nonumber
\end{equation}
\begin{equation}
=-\int \mathcal{D}[\psi^{\dagger},\psi] \ e^{\oint d\xi \  \psi^{\dagger}(\dot{\psi}+\dot{x}^{\mu}A_{\mu}\psi)} \bigg(\oint d\xi \ (\delta_1 x^{\mu}\delta_2 x^{\alpha}\dot{x}^{\nu}\psi^{\dagger}A_{\alpha}F_{\mu\nu}\psi +\nonumber
\end{equation}
\begin{equation}
\oint d\xi \  \delta_1 x^{\mu}\dot{x}^{\nu}\psi^{\dagger}F_{\mu\nu}\psi\oint d\xi' \ \delta_2x^{\alpha}\dot{x}^{\beta}\psi^{\dagger}A_{\alpha}A_{\beta}\psi\bigg) 
\end{equation}
and 
\begin{equation}
\int \mathcal{D}[\psi^{\dagger},\psi] \ e^{\oint d\xi \  \psi^{\dagger}(\dot{\psi}+\dot{x}^{\mu}A_{\mu}\psi)}  \oint d\xi \ \delta_1 x^{\mu}\dot{x}^{\nu}\psi^{\dagger}F_{\mu\nu}\psi\oint d\xi' \ \delta_2 x^{\alpha}\dot{\psi^{\dagger}}A_{\alpha}\psi  
\end{equation}
\begin{equation}
=\int \mathcal{D}[\psi^{\dagger},\psi] \ e^{\oint d\xi \  \psi^{\dagger}(\dot{\psi}+\dot{x}^{\mu}A_{\mu}\psi)}  \bigg(\oint d\xi \ (\delta_1 x^{\mu}\delta_2 x^{\alpha}\dot{x}^{\nu}\psi^{\dagger}F_{\mu\nu}A_{\alpha}\psi  \nonumber
\end{equation}
\begin{equation}
 +\oint d\xi \  \delta_1 x^{\mu}\dot{x}^{\nu}\psi^{\dagger}F_{\mu\nu}\psi\oint d\xi' \ \delta_2x^{\alpha}\dot{x}^{\beta}\psi^{\dagger}A_{\beta}A_{\alpha}\psi\bigg) \ .
\end{equation}
These allow us to write the second variation of $W_{\psi}$ as
\begin{equation}
\delta_2\delta_1W_{\psi}=\int \mathcal{D}[\psi^{\dagger},\psi] \ e^{\oint d\xi \  \psi^{\dagger}(\dot{\psi}+\dot{x}^{\mu}A_{\mu}\psi)}\bigg(\oint d\xi \ \psi^{\dagger}(\delta_1 x^{\mu}\delta_2 \dot{x}^{\nu}F_{\mu\nu}+\delta_1 x^{\mu} \delta_2 x^{\alpha}\dot{x}^{\nu}D_{\alpha}F_{\mu\nu})\psi \nonumber
\end{equation}
\begin{equation}
+\oint d\xi' \ \psi^{\dagger}\dot{x}^{\beta}\delta_2 x^{\alpha}F_{\alpha\beta}\psi     \oint d\xi \ \psi^{\dagger}\delta_1 x^{\mu}\dot{x}^{\nu}F_{\mu\nu}\psi\bigg).
\end{equation}

The Laplacian on loop space, $\Delta(\xi)$, acts to pick out the singular piece (also the first term vanishes since $F^{\mu}_{ \ \mu}=0$). Applying (\ref{area}) to $W_{\psi}$, using the above relations, we find
\begin{equation}
\Delta(\xi)W_{\psi}=\int \mathcal{D}[\psi^{\dagger},\psi] \ e^{\oint d\xi \  \psi^{\dagger}(\dot{\psi}+\dot{x}^{\mu}A_{\mu}\psi)}  \  \psi^{\dagger} \dot{x}^{\nu}D^{\mu}F_{\mu\nu}\psi|_{\xi}. \label{delta W}
\end{equation}
We can integrate out the gauge field by noticing that the integrand is proportional to the equations of motion for the Yang-Mills action. 
Under a variation of the gauge field, the functional integral
\begin{equation}
\int \mathcal{D}A \ e^{-S_{YM}} \ W_{\psi}
\end{equation}
changes as
\begin{equation}
0=\int \mathcal{D}A \ e^{-S_{YM}} \ \bigg(\int d^4x \ \frac{1}{q^2}\delta A^{\nu}D^{\mu}F_{\mu\nu}+\oint d\xi \ \psi^{\dagger}dx_{\nu}\delta A^{\nu}\psi\bigg).
\end{equation}
To compare this with the integrand of (\ref{delta W}), we choose the specific variation $\delta A^{\mu  A}=\delta^4(x-x(\xi'))\dot{x}^{\mu} \ \psi^{\dagger}\tau^A\psi$.
With this, we are finally lead to
\begin{equation}
 \Delta(\xi)\langle\,W_{\psi}\,\rangle= \nonumber
\end{equation}
\begin{equation}
-q^2 \int \mathcal{D}[\psi^{\dagger},\psi] \ \langle e^{\oint d\xi \  \psi^{\dagger}(\dot{\psi}+\dot x^{\mu}A_{\mu}\psi)} \rangle\oint \ \psi^{\dagger}\tau^A\psi \dot{x}^{\mu}|_{\xi}\delta^4(x(\xi)-x(\xi'))\psi^{\dagger}\tau^A\psi dx_{\mu}|_{\xi'}.
\end{equation}
This formula is the equivalent of (\ref{mandelstam}) for our generalised Wilson loop $W_\psi$.

\section{Supersymmetric Theory}
We can apply this procedure to the more applicable case of a non-Abelian gauge field coupled to fermions as in QCD described in the first-quantised worldline formalism.  To do this we give the loop spin degrees of freedom, which, in the worldline formalism corresponds to introducing a superpartner, $\eta^{\mu}$, for the co-ordinates, $x^{\mu}$. The super-Wilson loop is then
\begin{equation}
W_s\equiv \text{tr}\bigg( \ \mathcal{P} \ \text{exp}\bigg(\oint d\xi \bigg(\dot{x}^{\mu}A_{\mu}-\frac{\sqrt{h}}{2}\eta^{\mu}\eta^{\nu}F_{\mu\nu}\bigg)\bigg)\bigg)\,.
\end{equation}
The super-Wilson loop is invariant under the $N=1$ worldline supersymmetry transformations, parametrised by the Grassmann-odd function $\epsilon(\xi)$,
\begin{equation}
\delta x^{\mu}=-\epsilon\sqrt{h}\eta^{\mu} , \ \ \ \ \delta \eta^{\mu}=\epsilon\dot{x}^{\mu}    .
\end{equation}
The loop equations are most easily obtained by appealing to the superspace formalism. We introduce the anti-commuting variable, $\theta$, as the superpartner of the worldline parameter, $\xi$, related to it by the supersymmetry transformations, $\delta \theta=-\tilde{\epsilon}$ and $\delta \xi=\tilde{\epsilon}\theta$. We also introduce the 1 dimensional superfield, $\mathbb{X}^{\mu}$, which can be expanded in powers of $\theta$ as
\begin{equation}
\mathbb{X}^{\mu}=x^{\mu}+ih^{1/4}\theta\eta^{\mu}.
\end{equation}
along with the superderivative $\mathcal{D}\equiv \partial_{\theta}+\theta\partial_{\xi}$.
The super-Wilson loop can then be written as
\begin{equation}
W_s=\text{tr}\bigg( \ \mathcal{P} \ \text{exp}\bigg(\oint d\xi d\theta \ \mathcal{D} \mathbb{X}^{\mu}A_{\mu}(\mathbb{X}) \bigg)\bigg).
\end{equation}
To see the equivalence, one needs the superspace analogue of path-ordering \cite{Andreev}.

The Mandelstam equation for the super-Wilson loop is then
\begin{equation}
\frac{1}{Z_0}\int \mathcal{D}A \ e^{-S_{YM}} \ \Delta_s(\xi)W_s= \nonumber
\end{equation}
\begin{equation}
-q^2 \ \text{tr} \ \bigg( \mathcal{P} \ \int d\xi'd\theta' d\theta \ \tau^A\mathcal{D}\mathbb{X}^{\mu}|_{\xi',\theta'} \ \delta^4(\mathbb{X}(\xi')-\mathbb{X}(\xi)) \ \tau^A\mathcal{D}\mathbb{X}_{\mu}|_{\xi,\theta}  \ e^{ \int d\xi d\theta \ \mathcal{D}\mathbb{X}^{\nu}A_{\nu}(\mathbb{X})}\bigg)
\end{equation}
where
\begin{equation}
\tilde{\Delta}(\xi)\equiv \lim_{\epsilon\rightarrow 0}\int^{\epsilon}_{-\epsilon} d\xi'd\theta d\theta' \ \frac{\delta^2}{\delta \mathbb{X}^{\mu}((\xi,\theta)+(\xi',\theta')/2)\delta \mathbb{X}_{\mu}((\xi,\theta)-(\xi',\theta')/2)} \label{susy lap}
\end{equation}
generalises the Laplacian on loop space.

The path-ordering of the super-Wilson loop can be produced in an analogous manor to that of the bosonic case. The loop variable we will consider this time is a straightforward generalisation of the bosonic loop
\begin{equation}
W_{\psi s}=\int \mathcal{D}[\psi^{\dagger},\psi] \ \text{exp}\bigg(\int d\xi \ \psi^{\dagger}\bigg(\frac{d}{d\xi}+\dot{x}^{\mu} A_{\mu}-\frac{\sqrt{h}}{2} \ \eta^{\mu}F_{\mu\nu}\eta^{\nu}\bigg)\psi\bigg). \label{super wil psi}
\end{equation}
We are unable to write this as a superspace integral since $\psi^{\dagger}$ and $\psi$ don't have superpartners. We can introduce these though by considering the integral
\begin{equation}
\int d\xi \ \left(\psi^{\dagger}\left(\frac{d}{d\xi}+\dot{x}^{\mu} A_{\mu}-\sqrt{h} \ \eta^{\mu}\partial_{\mu}A_{\nu}\eta^{\nu}\right)\psi+\sqrt{h} \ (z^{\dagger}\eta^{\mu}A_{\mu}\psi+\psi^{\dagger}\eta^{\mu}A_{\mu}z+\tilde{z}z)\right).
\end{equation}
Integrating out $z^{\dagger}$ and $z$ reduces this to the argument of the exponential in the super-Wilson loop. Not only does this have the benefit of being linear in the gauge field, the integral is invariant under $\delta z^{\dagger}=\tilde{\epsilon}\psi^{\dagger}$ and $\delta \psi^{\dagger}=-\tilde{\epsilon}$, with similar transformations for $z$ and $\psi$. We can then introduce the superfields $\Gamma$ and $\Gamma^{\dagger}$ defined by the respective expansions
\begin{equation}
\Gamma\equiv \psi+ih^{1/4}\theta \ z
\end{equation}
\begin{equation}
\Gamma^{\dagger}\equiv \psi^{\dagger}+ih^{1/4}\theta \ z^{\dagger}
\end{equation}
so that (\ref{super wil psi}) can be written as
\begin{equation}
W_{\Gamma}\equiv\int \mathcal{D}[\Gamma^{\dagger},\Gamma] \ \text{exp}\bigg(\int d\xi d\theta \ \Gamma^{\dagger}(\mathcal{D}+\mathcal{D}\mathbb{X}^{\mu} A_{\mu}(\mathbb{X}))\Gamma\bigg).
\end{equation}
We can now go ahead and study the change in the loop variable under a variation of $\mathbb{X}$. Using the same method from section 2, we find for the first variation
\begin{equation}
\delta_{\mathbb{X}}W_{\Gamma}=\int \mathcal{D}[\Gamma^{\dagger},\Gamma] \ e^{\int d\xi d\theta \ \Gamma^{\dagger}(\mathcal{D}+\mathcal{D}\mathbb{X}^{\mu} A_{\mu}(\mathbb{X}))} \ \int d\xi d\theta \   \mathcal{D}\mathbb{X}^{\mu}\delta\mathbb{X}^{\nu} \ \tilde{\Gamma}F_{\mu\nu}(\mathbb{X})\Gamma
\end{equation}
and for the second variation
\begin{equation}
\delta_2\delta_1W_{\Gamma}=\int \mathcal{D}[\Gamma^{\dagger},\Gamma] \  e^{\int d\xi d\theta \ \Gamma^{\dagger}(\mathcal{D}+\mathcal{D}\mathbb{X}^{\mu} A_{\mu}(\mathbb{X}))} \bigg( \int d\xi d\theta \   \delta_2(\mathcal{D}\mathbb{X}^{\mu})\delta_1\mathbb{X}^{\nu} \ \tilde{\Gamma}F_{\mu\nu}(\mathbb{X})\Gamma \nonumber
\end{equation}
\begin{equation}
+ \int d\xi d\theta \ \mathcal{D}\mathbb{X}^{\mu}\delta_1\mathbb{X}^{\nu}\delta_2\mathbb{X}^{\alpha} \ \tilde{\Gamma} \ D_{\alpha}F_{\mu\nu}(\mathbb{X})\Gamma  \nonumber
\end{equation}
\begin{equation}
- \int d\xi_1 d\theta_1 \   \mathcal{D}\mathbb{X}^{\mu}\delta_1\mathbb{X}^{\nu} \ \tilde{\Gamma}F_{\mu\nu}(\mathbb{X})\Gamma\int d\xi_2 d\theta_2 \   \mathcal{D}\mathbb{X}^{\beta}\delta_2\mathbb{X}^{\alpha} \ \tilde{\Gamma}F_{\alpha\beta}(\mathbb{X})\Gamma \bigg).
\end{equation}
Using this result we are able to compute the Laplacian (\ref{susy lap}) applied to $W_{\Gamma}$,  which picks out the second line exclusively
\begin{equation}
\tilde{\Delta}(\xi)W_{\Gamma}= \int \mathcal{D}[\Gamma^{\dagger},\Gamma] \ e^{\int d\xi d\theta \ \Gamma^{\dagger}(\mathcal{D}+\mathcal{D}\mathbb{X}^{\mu} A_{\mu}(\mathbb{X}))} \ \int d\theta \ \mathcal{D}\mathbb{X}^{\nu} \ \tilde{\Gamma} \ D^{\mu}F_{\mu\nu}(\mathbb{X}) \ \Gamma |_{\xi}. \label{area sus}
\end{equation}
We are using the definition of functional differentiation
\begin{equation}
\frac{\delta \mathbb{X}^{\mu}(\xi,\theta)}{\delta \mathbb{X}^{\nu}(\xi',\theta')}=\delta^{\mu}_{ \ \nu}\delta(\xi-\xi')(\theta-\theta').
\end{equation}
The gauge field can be integrated out as in the bosonic case. In this case the appropriate functional integral to consider is
\begin{equation}
\int \mathcal{D}A \ e^{-S_{YM}+\int d\xi d\theta \ \Gamma^{\dagger}(\mathcal{D}+\mathcal{D}\mathbb{X}^{\mu} A_{\mu}(\mathbb{X}))}
\end{equation} 
where $S_{YM}$ is still given by (\ref{maxwell}). This means we must compare $A_{\mu}(x)$ with the superfield $A_{\mu}(\mathbb{X})$. Consider the change in this integral under a variation of the gauge field, we find
\begin{equation}
0=\int \mathcal{D}A \ e^{-S_{YM}+\int d\xi d\theta \ \Gamma^{\dagger}(\mathcal{D}+\mathcal{D}\mathbb{X}^{\mu} A_{\mu}(\mathbb{X}))} \bigg(\int d^4x \ \frac{1}{q^2}\delta A^{\nu}(x)D^{\mu}F_{\mu\nu}(x)+ \nonumber
\end{equation}
\begin{equation}
\int d\xi d\theta \ \Gamma^{\dagger}\mathcal{D}\mathbb{X}^{\nu}\delta A_{\nu}(\mathbb{X})\Gamma\bigg)
\end{equation}
where
\begin{equation}
\delta A_{\nu}(\mathbb{X})=\int d^4x' \ \delta^4(x'-\mathbb{X})A_{\nu}(x')
\end{equation}
To make the connection with (\ref{area sus}), we require 
\begin{equation}
\delta A^{\nu A}(x)=q^2 \ \mathcal{D}\mathbb{X}^{'\nu} \ \Gamma^{\dagger}\tau^A\Gamma \ \delta^4(x-\mathbb{X}')
\end{equation}
so that 
\begin{equation}
\delta A_{\nu}(\mathbb{X})=q^2\int d^4x'' \ \delta^4(x''-\mathbb{X})\delta^4(x''-\mathbb{X}')\mathcal{D}\mathbb{X}'_{\nu} \  \Gamma^{\dagger}\tau^A\Gamma \nonumber
\end{equation}
\begin{equation}
=q^2 \ \delta^4(\mathbb{X}-\mathbb{X}')\mathcal{D}\mathbb{X}'_{\nu} \ \Gamma^{\dagger}\tau^A\Gamma.
\end{equation}
Now, we can use this to write
\begin{equation}
 \tilde{\Delta}(\xi)\,W_{\Gamma}\,\rangle=-q^2 \langle\,\int \mathcal{D}[\Gamma^{\dagger},\Gamma] \ \langle e^{\int d\xi d\theta \ \Gamma^{\dagger}(\mathcal{D}+\mathcal{D}\mathbb{X}^{\mu} A_{\mu}(\mathbb{X}))}\rangle \times \nonumber
\end{equation}
\begin{equation}
 \int  d\theta \int d\xi'd\theta' \ (\mathcal{D}\mathbb{X}^{\nu}  \ \tilde{\Gamma} \tau^A \ \Gamma) |_{\xi,\theta} \ \delta^4(\mathbb{X}-\mathbb{X}') \ (\mathcal{D}\mathbb{X}_{\nu} \ \Gamma^{\dagger}\tau^A\Gamma)|_{\xi',\theta'}.
\end{equation}
This concludes the derivation of the Mandelstam formula for the supersymmetric version of our generalised Wilson loop.
\section{Conclusion}
We have derived Mandelstam formulae for two generalisations of the Wilson loop. The Mandelstam formula for the usual Wilson loop is a key step in deriving the loop equations of Migdal and Makeenko which capture the dynamics of quantum gauge theories. The generalisations of the Wilson loop we considered are potentially useful because they can be used to encode the representations and helicities that appear in the Standard Model in a simple way. They also arise as the boundary theories of the string model considered in \cite{Curry}, which gives an attempt at a reformulation of Yang-Mills theory based on a generalisation of the model in \cite{Edwards:2014xfa}. This attempt is as yet incomplete, but in pursuing this search further it is necessary to be able to compare the dynamics of the string model with those of Yang-Mills theory and we expect that the loop equations will provide the appropriate tool to do this.

\acknowledgments
 CC is grateful to STFC for a studentship.
This research is also supported by the Marie Curie network GATIS 
of the European Union's Seventh Framework Programme FP7/2007-2013/ under REA Grant Agreement No 317089.

\end{document}